%% file: main.tex
\documentclass[lettersize,journal]{IEEEtran}
    \usepackage{booktabs}
    \usepackage{diagbox}
    \usepackage{makecell}
    \usepackage{array}
    \usepackage{graphicx,amssymb,amsmath}
    \usepackage{multicol}
    \usepackage[noadjust]{cite} 
    \usepackage{setspace}
    \usepackage{subcaption} 
    \usepackage{graphicx}
    \usepackage{float}
    \usepackage {url}
    \usepackage{stfloats}
    \usepackage{amsthm,pifont}
    \usepackage{cases,subeqnarray}
    \usepackage{bm,multirow,bigstrut}
    \usepackage{amsmath, amsthm, amssymb}
    \usepackage{textcomp}
    \usepackage{latexsym,bm}
    \usepackage{booktabs}
    \usepackage{mathtools}
    \usepackage{dsfont}
    \usepackage{extarrows}
    \usepackage{epsfig}
    \usepackage{epsfig}
    \usepackage{epstopdf}
    \usepackage[table]{xcolor}
    \usepackage{colortbl} 
    \usepackage{multirow, hhline}
    \usepackage{xcolor}
    \usepackage{array} 
    \usepackage{algorithmicx,algorithm}
    \usepackage{hyperref}

    \theoremstyle{plain}

    \theoremstyle{plain}

    \usepackage{amsmath}

    \IEEEoverridecommandlockouts
    
\begin{document}
    \title{Generative AI-enabled Wireless Communications for Robust Low-Altitude Economy Networking}
    
    \author{
    Changyuan Zhao, Jiacheng Wang, Ruichen Zhang, Dusit Niyato,~\IEEEmembership{Fellow,~IEEE}, Geng Sun, Hongyang Du,\\
    Dong In Kim,~\IEEEmembership{Fellow,~IEEE}, Abbas Jamalipour,~\IEEEmembership{Fellow,~IEEE}
    \thanks{C. Zhao is with the College of Computing and Data Science, Nanyang Technological University, Singapore, and CNRS@CREATE, 1 Create Way, 08-01 Create Tower, Singapore 138602 (e-mail: zhao0441@e.ntu.edu.sg).}
    \thanks{J.~Wang, R. Zhang, and D. Niyato are with the School of Computer Science and Engineering, Nanyang Technological University, Singapore (e-mail: jiacheng.wang@ntu.edu.sg, ruichen.zhang@ntu.edu.sg, dniyato@ntu.edu.sg).}
        \thanks{G. Sun is with College of Computer Science and Technology, Jilin University, China 130012, (e-mail: sungeng@jlu.edu.cn).}
    \thanks{H. Du is with the Department of Electrical and Electronic Engineering,
University of Hong Kong, Hong Kong (e-mail: duhy@eee.hku.hk).}
    \thanks{D. I. Kim is with the Department of Electrical and Computer Engineering, Sungkyunkwan University, Suwon 16419, South Korea (e-mail:dongin@skku.edu).}
    \thanks{A. Jamalipour is with the School of Electrical and Computer Engineering,
University of Sydney, Australia (e-mail: a.jamalipour@ieee.org).}
    }
\maketitle
    \begin{abstract}

Low-Altitude Economy Networks (LAENets) have emerged as significant enablers of social activities, offering low-altitude services such as the transportation of packages, groceries, and medical supplies. Owing to their control mechanisms and ever-changing operational factors, LAENets are inherently more complex and vulnerable to security threats than traditional terrestrial networks. As applications of LAENet continue to expand, the robustness of these systems becomes crucial. In this paper, we propose a generative artificial intelligence (GenAI) optimization framework that tackles robustness challenges in LAENets. We conduct a systematic analysis of robustness requirements for LAENets, complemented by a comprehensive review of robust Quality of Service (QoS) metrics from the wireless physical layer perspective. We then investigate existing GenAI-enabled approaches for robustness enhancement. This leads to our proposal of a novel diffusion-based optimization framework with a Mixture of Experts (MoE)-transformer actor network. In the robust beamforming case study, the proposed framework demonstrates its effectiveness by optimizing beamforming under uncertainties, achieving a more than 15\% increase over four learning baselines in the worst-case achievable secrecy rate. These findings highlight the significant potential of GenAI in strengthening LAENet robustness.
    
\end{abstract}
    \IEEEpeerreviewmaketitle

    \input{intro}

    \input{section2}

    \input{section3}
    \input{section5_new}
    \input{con}
    \bibliographystyle{IEEEtran}
    \bibliography{IEEEabrv,Ref}









\end{document}

%% file: intro.tex
\section{Introduction}






Recently, advancements in wireless communications 
and manufacturing technologies have propelled the development of drones and other aerial devices, greatly expanding their applications in low-altitude airspace.
Enabled by these technologies, the Low-Altitude Economy (LAE) has emerged as a significant sector focusing on social opportunities within low-altitude airspace (defined as elevations up to 1,000 meters).
The LAE aims to harness low-flying equipment for various social activities, including air traffic management, transportation, and surveillance, to organize the Low-Altitude Economy Networks (LAENet). 
These networks play an increasingly vital role by enabling essential services such as communication, computation, storage, and sensing \cite{zheng2025uav}.
According to the Civil Aviation Administration of China, the value of China's low-altitude airspace economy exceeded CNY 500 billion in 2023 and is projected to reach CNY 3.5 trillion by 2035\footnote{https://www.swissre.com/institute/research/topics-and-risk-dialogues/china/insuring-low-altitude-airspace-economy-china.html}.
Moreover, 
the U.S. Federal Aviation Administration is set to finalize pilot training and certification rules for air taxis by 2024. 
A well-known research company, Mordor Intelligence, predicts that the urban air mobility market will reach approximately \$4.04 billion by 2029,
demonstrating its significant social potential\footnote{https://www.moomoo.com/news/post/44869062/faa-finalized-key-rules-usa-s-low-altitude-economy-leader}.



However, 
the unique low-altitude environment and the absence of unified management regulations have introduced several uncertainties to the design of LAENet.
These uncertainties, including variable weather conditions, air traffic, user demand, and emergency situations, significantly heighten the requirement for LAENet devices to complete tasks in dynamic environments  \cite{zheng2025uav}.
Therefore, robustness has been recognized as a Key Performance Indicator (KPI) that goes beyond efficiency (speed) and effectiveness (accuracy) for LAENet. 
\textit{Robustness refers to the ability to maintain the qualities of service (QoS) of LAENets despite the influence of uncertainties} \cite{mozaffari2019tutorial}. This recognition guides the evolution of LAENets, enabling innovation and commercial activity.
For example, 
the tech giant Uber's drone delivery utilizes advanced artificial intelligence algorithms, Global Positioning System (GPS) systems, and sensors to avoid obstacles under different weather conditions, ensuring maximum delivery safety\footnote{https://karlobag.eu/en/technology/uber-launches-innovative-drone-food-delivery-service-in-urban-areas-using-high-technology-and-safety-measures-0tvcw}.
However, due to the complexity of the airspace environment, the high degrees of freedom, and the strong dynamics of aerial devices, building a robust wireless network platform for LAE applications is challenging.

Enhancing the robustness of the wireless physical layer requires addressing uncertainty as a critical issue.
However, both numerical algorithms and traditional learning methods face several challenges, including heavy dependence on training data, and susceptibility to unknown environments, which hinder their effectiveness in handling uncertainty.
Generative artificial intelligence (GenAI) is an emerging technology that works to analyze data distribution.
GenAI has demonstrated strong capabilities in enhancing wireless communications, including physical and network layers, through cross-dimensional data processing and generation \cite{zhao2024generative}. 
Importantly, recent research highlights the potential of GenAI to capture the underlying distribution of optimal strategies, enhancing network optimizations \cite{du2024enhancing}.
Based on the achievements in wireless communications, there are several potential
advantages of using GenAI to eliminate uncertainties and
construct robust LAENets.
GenAI can generate additional signal data, which is difficult to obtain in extreme scenarios, to handle diverse interference patterns. Additionally, GenAI can generate clean data, thereby removing the noise uncertainty in signals~\cite{zhao2024generative}. 
Moreover, GenAI enables optimization by modeling the distribution of strategies under uncertainties, allowing it to generate robust and improved solutions.



GenAI models with transformer architectures, such as Large Language Models (LLMs) and Deepseek \cite{liu2024deepseek}, effectively capture long-range dependencies and contextual information. They are robust against noisy inputs and distribution shifts, making them useful for semantic communication, adaptive coding, and efficient spectrum utilization in wireless networks~\cite{liu2024generative}. Mixture of Experts (MoE) architectures enhance computational efficiency by activating relevant expert pathways, improving scalability and reducing errors by adapting to changing channel conditions \cite{zhao2025enhancing}.

Motivated by these analyses, this paper aims to investigate the utilization of GenAI to develop robust LAENets. The main contributions are summarized as follows:
\begin{itemize}
    \item We provide an overview of LAENets and their applications, offering a detailed analysis of the robustness within LAENet. Furthermore, we introduce various GenAI models within their applications and benefits in wireless communications toward LAENets.
    \item From the perspective of the physical layer, we explore how GenAI enhances the robustness of wireless communications. Moreover, we present comprehensive robustness metrics and discuss the benefits of implementing GenAI models to support the construction of LAENets.

    \item To systematically tackle robustness challenges, we incorporate robustness requirements into an optimization problem. We consider different levels of robustness through different optimization paradigms, including stochastic optimization, chance-constrained optimization, and robust optimization.
    
    \item We propose an MoE-transformer actor network integrated with the diffusion-based optimization framework in LAENets. Through a case study, we illustrate how GenAI can facilitate the development of robust beamforming for LAENets.
\end{itemize}

%% file: section2.tex
\section{The Overview of Robust LAENet}

In this section, we first introduce the LAENet and its related applications. Further, we discuss the robust issues and challenges in the LAENets. 


\begin{figure*}[htp]
    \centering
    \includegraphics[width= 0.85\linewidth]{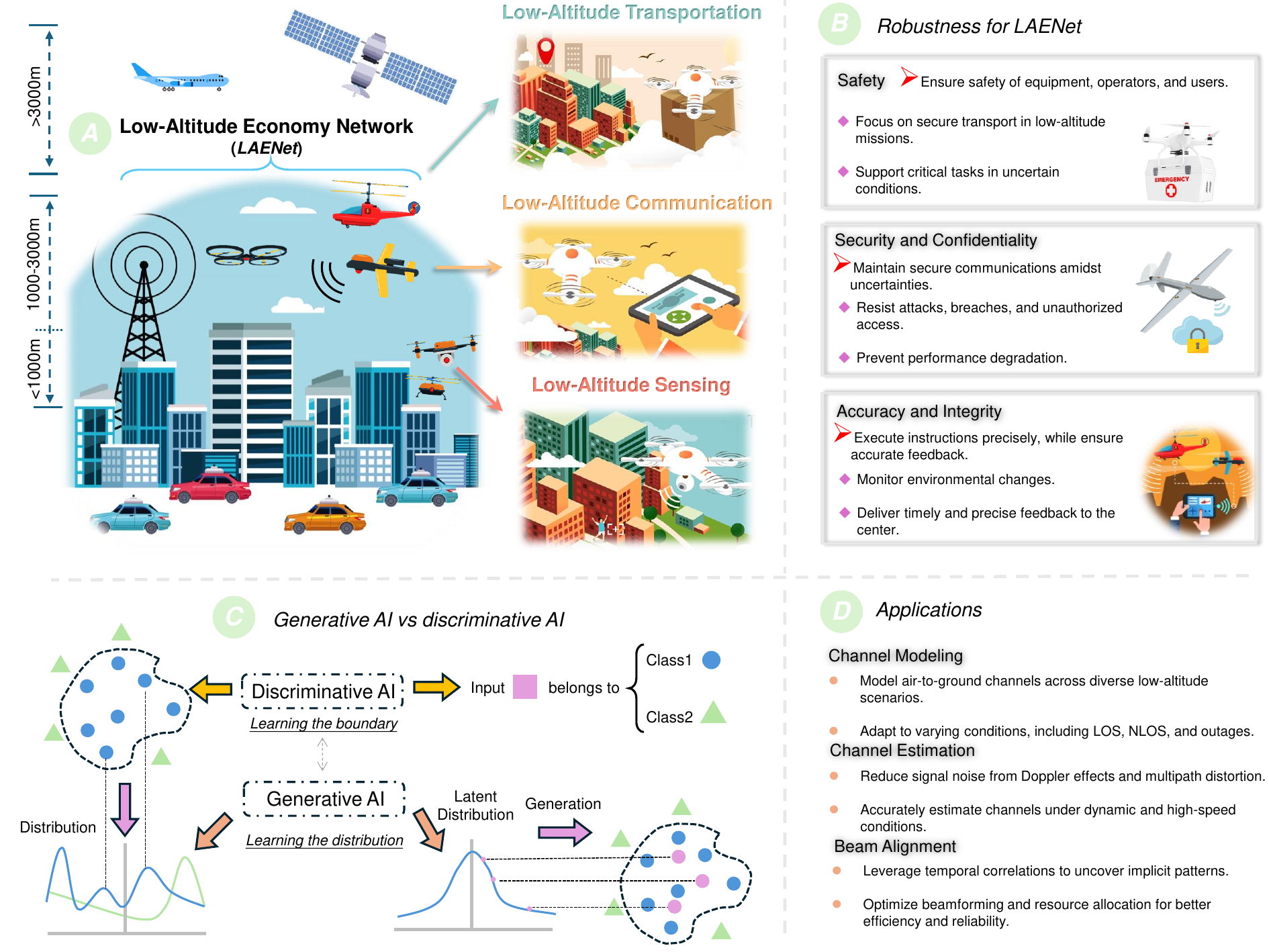}
    \caption{An overview of GenAI-enabled robust LAENet. \textit{Part A} illustrates the architecture of LAENet from transportation, communication, and sensing aspects.
    \textit{Part B} introduces three aspects of robustness in LAENets.
    \textit{Part C} presents a comparison between GenAI and discriminative AI.
    \textit{Part D} demonstrates important GAI-enabled applications, including channel modeling, channel estimation, and beam alignment.
    }
    \label{fig:frame}
\end{figure*}

\subsection{Conceptions and Applications of LAENet}

The LAE relies on the cooperation of edge equipment and low-altitude communication networks to facilitate social activities, which covers three primary aspects (shown in Fig. \ref{fig:frame} \textit{Part A}):

\begin{itemize}
\item \textbf{Low-Altitude Transportation}:
One of the most prominent applications of low-altitude transportation is drone-based delivery. 
Companies including Amazon, Walmart, and Wing are pioneering the use of drones to transport packages, groceries, and medical supplies.

\item \textbf{Low-Altitude Communication}:
LAENets can address the limitations of traditional communication systems in remote and rural areas.
For instance, Telia Company has deployed drones equipped with portable mobile base stations to enable remote control of forestry machinery over 5G networks.






    \item \textbf{Low-Altitude Sensing}:
Drones are invaluable tools for monitoring environmental changes. Drone enables high-resolution mapping and data collection for tracking deforestation, glacier melting, and wildlife habitats. Their ability to access hard-to-reach areas ensures comprehensive coverage, supporting effective conservation strategies.


\end{itemize}

\subsection{The Significance of Robust LAENet}


In contrast to traditional terrestrial networks, LAENets are distinguished by automatic control mechanisms and constantly shifting operational factors, making them more complex and vulnerable. 
As applications such as aerial deliveries grow, maintaining robustness becomes crucial. 
\textit{Robustness refers to the ability to maintain the QoS of LAENet despite the influence of uncertainties} \cite{mozaffari2019tutorial}.
Many LAENet use cases are mission-critical, where even small security compromises can significantly impact QoS. Therefore, ensuring robust design and operation in these networks is essential, focusing on three core elements as shown in Fig. \ref{fig:frame} \textit{Part B}.

\begin{itemize}

  \item \textbf{Safety}: 
Safety in LAENets involves ensuring the protection of equipment, operators, and service recipients during operations with low-altitude devices. 
Ensuring QoS of safety means that safety measures must consistently perform effectively, even when faced with challenges such as adverse weather, signal interference, or unexpected obstacles.
        

    \item \textbf{Security and Confidentiality}: 
    Security and confidentiality concerns LAENet's ability to maintain secure and reliable low-altitude communications while ensuring QoS amidst various uncertainties. This involves resisting external attacks, protecting against data breaches, and preventing unauthorized access, all without compromising the performance required for different applications. 

        \item \textbf{Accuracy and Integrity}: 
    Accuracy and integrity encompass LAENet's ability to transmit and execute instructions precisely and provide accurate feedback on the operational status of devices. 
    This ensures that QoS is maintained, even with unpredictable environmental factors or uncertainties in data transmission.


\end{itemize}

Therefore, designing and operating robust LAENet is imperative to ensure reliable, secure, and high quality service across diverse applications.

\subsection{Robustness Considerations for LAENet}

\begin{figure*}[htp]
    \centering
    \includegraphics[width= 0.95\linewidth]{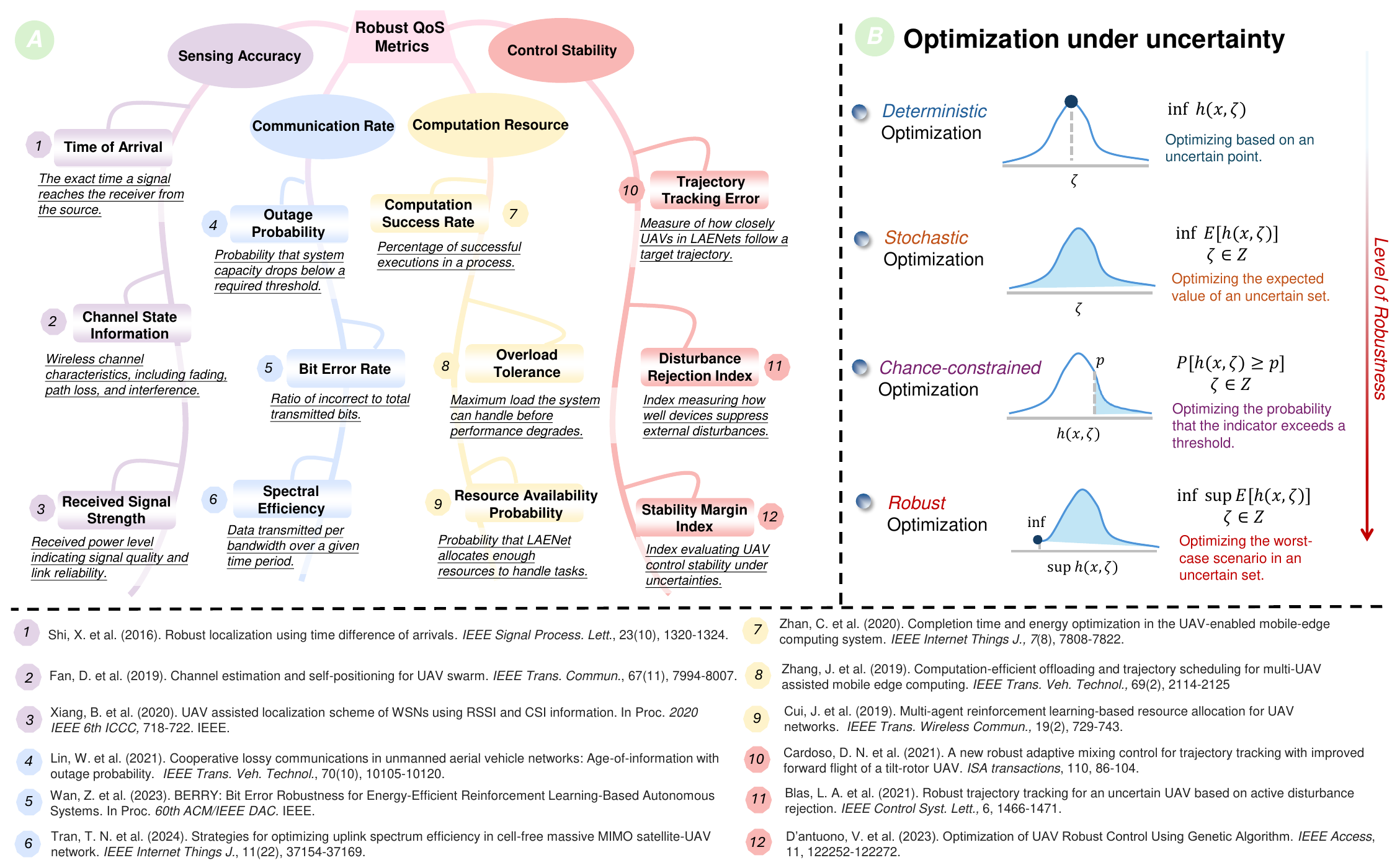}
    \caption{An illustration of robustness considerations in LAENet. \textit{Part A} presents several QoS metrics. \textit{Part B} introduces 4 uncertainty optimizations.}
    \label{fig:robust_metrics}
\end{figure*}

To effectively evaluate and enhance the robustness of LAENet, it is necessary to introduce robust QoS metrics in various applications. 
We first map key LAENet applications to suitable QoS indicators, transforming them into optimization problems under uncertainty, as illustrated in Fig. \ref{fig:robust_metrics}.
\begin{itemize}
\item \textbf{Sensing Accuracy Indicators}:
These indicators
evaluate spatial, frequency, and power domain features, enabling more reliable and resilient environmental perception to enhance the robustness, including time of arrival, channel state information, and received signal strength representing the performance of LAENets.
\item \textbf{Communication Rate Indicators}:
The robustness of the communication rate can be improved by minimizing transmission failures and errors while maximizing data throughput across varying channel conditions, which can be measured by outage probability, bit error rate (BER), spectral efficiency, etc.
\item \textbf{Computation Resource Indicators}:
These indicators enhance data transmission efficiency, reduce signaling overhead, and accelerate processing in dynamic network conditions, as measured by computation success rate, overload tolerance, and resource availability probability.
\item \textbf{Control Stability Indicators}:
Control stability influences the speed, trajectory, and state of the device, significantly impacting channel quality in LAENets due to distance, fading, and shadowing.
Control stability can be achieved by optimizing trajectory tracking accuracy, disturbance rejection, and the stability margin index to maintain stability in dynamic environments.


\end{itemize}

Besides robust indicators, the effective adjustment and improvement of these indicators are typically realized through optimization methods. However, recognizing the varying nature of uncertainties inherent to LAENet operations, it is necessary to employ appropriate optimization frameworks to systematically address and manage these uncertainties.
We introduce several optimization frameworks that offer increasing levels of robustness, including deterministic, stochastic, chance-constrained, and robust optimization \cite{ben2009robust}, as shown in Fig. \ref{fig:robust_metrics} \textit{Part B}.



Deterministic optimization, which assumes a predictable environment, maximizes performance indicators under fixed conditions. It is suitable for scenarios where uncertainties are negligible. Conversely, stochastic optimization explicitly accounts for uncertainties by optimizing expected performance across a range of scenarios, making it effective in moderately uncertain environments.
For scenarios with strict QoS requirements but limited error tolerance, chance-constrained optimization is employed. It ensures predefined requirements are met with a probabilistic guarantee, balancing performance and risk by controlling acceptable levels of constraint violation~\cite{ben2009robust}.

Robust optimization provides the highest level of reliability by optimizing worst-case performance, ensuring that LAENet systems remain operational under severe uncertainties or adversarial conditions. This method is particularly important in mission-critical applications where failure could result in significant consequences~\cite{ben2009robust}.

By aligning each robustness indicator with an appropriate optimization level, these optimization frameworks systematically enhance LAENet robustness, quantitatively improving critical QoS metrics under environmental uncertainties.

%% file: section3.tex
\section{GenAI-enabled Wireless Physical Layer for Robust LAENet}


In this section, we provide an overview of GenAI.
Then, we present some GenAI-enabled wireless physical layer communication methods and show how these methods improve the robust QoS metrics for LAENets.

\subsection{Overview of GenAI}



GenAI is a type of Artificial Intelligence (AI) that 
leverages unsupervised learning or self-supervised learning to generate new results that resemble the learned data, as depicted in Fig. \ref{fig:frame} \textit{Part C}. 
Typical generative AI models include Generative Adversarial Networks (GANs), Variational Autoencoders (VAEs), and diffusion models, LLM \cite{zhao2024generative}.
GenAI excels in capturing and modeling specific data distributions, which allows for more accurate and detailed data analysis.
For example, GenAI can detect subtle deviations that may indicate operational safety issues, enabling accurate threat detection. Additionally, it can generate synthetic data to simulate potential threats, such as spoofing attacks or signal interference in low-altitude systems, providing an adaptation mechanism~\cite{liu2024generative}.
These features can address robust issues encountered by LAENets, providing a solid foundation for leveraging GenAI models to enhance the robustness of LAENets in wireless physical layer communications.

\subsection{Comparison of GenAI and discriminative AI}

As shown in Fig. \ref{fig:frame} \textit{Part C}, GenAI can generate novel content by learning and mimicking the underlying data distributions. Distinct from traditional learning paradigms, such as discriminative AI, which are primarily designed to produce specific outcomes or decisions, 
GenAI excels in capturing and modeling specific data distributions, which allows for more accurate and detailed data analysis \cite{zhao2025enhancing}.
For instance, by learning the distributions of attacked and unattacked data, GenAI can analyze whether the input data conforms to these distributions, offering insights into potential anomalies and enhancing robustness, rather than merely making binary judgments about whether the input data has been attacked.

For optimization frameworks, discriminative AI methods traditionally employ deterministic optimization focusing primarily on decision boundaries. However, their performance degrades rapidly under high uncertainties \cite{zhao2025enhancing}.
In contrast, GenAI methods inherently model underlying data distributions, 
enabling a probabilistic representation of uncertainties that aligns naturally with stochastic, chance-constrained, and robust optimization paradigms.
This probabilistic foundation allows GenAI to offer superior capabilities for robust optimization~\cite{du2024enhancing}.
Additionally, GenAI addresses the limitations of traditional methods by generating additional synthetic data under extreme scenarios, removing noise uncertainties, and simulating realistic environmental conditions, thereby quantitatively enhancing robustness metrics such as BER, outage probability, and stability margins~\cite{zhao2024generative}.

\subsection{Applications}

\subsubsection{Channel Modeling}

Millimeter-wave (mmWave) technology is the primary choice in low-altitude communications, due to its substantial bandwidth and line-of-sight (LOS) capabilities. The design and evaluation of mmWave LAENets heavily depend on the availability of robust channel models in various low-altitude scenes. However, in aerial communications, this complexity is further compounded by factors such as UAV altitude, 3D orientation, and the height of surrounding buildings.
In \cite{xia2022generative}, the authors introduce a two-stage GenAI model for robust and versatile air-to-ground channel modeling. In the first stage, a fully connected neural network predicts the link state, determining whether the channel is in LOS, Non-Line of Sight (NLOS), or experiencing an outage. The second stage involves a conditional VAE that generates path parameters, including path losses, delays, and angles of arrival and departure.
Through VAE-generated path parameters, air-to-ground channel modeling can be adapted to various low-altitude environments, such as LOS, NLOS, and outage conditions.
With a more accurate channel model, the indicators, including path loss, packet delivery ratio, and fading stability, can be significantly improved.
Compared with the Third Generation Partnership Project (3GPP) channel model, 
the proposed approach outperforms 3GPP models and the refitted 3GPP model to effectively capture complex statistical relationships in the channel modeling data \cite{xia2022generative}. 
The proposed method reduces path loss in various cities by 9dB to 18dB compared to the standard model, which aligns more closely with actual values. The results demonstrate the accuracy of the GenAI method in channel modeling in various low-altitude environments.




\subsubsection{Channel Estimation}

In low-altitude transportation scenarios, 
Orthogonal Time Frequency Space (OTFS) can select the appropriate channel estimation technique to ensure robust communications.
However, traditional estimators, including least squares (LS), struggle to learn channel parameters in high-speed scenarios.
Motivated by the generalization capabilities of GenAI, the authors in \cite{gupta2024gance} propose a GAN-based channel estimation model, which learns from channel data rather than relying on strict mathematical models, allowing it to generalize to different channel environments.
The received signal is demodulated using an OTFS demodulator and then fed to the GAN block for channel estimation, where the GAN block is trained based on 10000 high-speed data sets with the receiver velocity varying from 100 to 500 km/h.
Leveraging the GAN block, the received signal becomes less susceptible to the Doppler effect and multi-path distortion during channel propagation. This enables more accurate channel estimation, which can enhance indicators, such as BER, packet error probability, and diversity gain.
Compared to traditional LS estimators, machine learning-based estimators, and deep learning-based estimators, the proposed GAN enhances BER performance by 70\%, 55\%, and 45\%, respectively. Additionally, it improves robust probability indicator outage probability by 40\%, 30\%, and 20\%, respectively \cite{gupta2024gance}. 

\subsubsection{Beam Alignment}




In LAENets, 
mmWave communication faces significant beam training overhead due to the mobility of a user equipment (UE) or a base stations (BS) in LAENets. 
Traditional beam alignment methods, such as exhaustive search, suffer from excessive overhead, as they lack closed-form expression.
The authors in \cite{hussain2021learning} propose a deep recurrent (DR-)VAE, designed to learn the dynamics of the strongest beam pair. 
The proposed DR-VAE extends the VAE framework to predict the probability of the strongest beam pair and beam training feedback, which then minimizes overhead and maximizes frame spectral efficiency.
The VAE framework handles temporally correlated observations, 
which facilitates a deeper analysis of the implicit relationships between action-observation sequences. These relationships can optimize resource allocation effectively. Robust QoS metrics, such as spectral efficiency and link reliability, can be enhanced in capacity and reliability.
Numerical evaluations employing 3D simulated beamforming at both the BS and UE demonstrate that integrating DR-VAE with beam alignment strategies reduces the Kullback-Leibler divergence between the ground truth and the learned model by approximately 95\%. The proposed method outperforms the Baum-Welch algorithm, achieving an improvement in spectral efficiency of about 10\%~\cite{hussain2021learning}.
\subsection{Lessons Learned}

Building on prior research, GenAI has shown significant potential in physical layer wireless communications, especially in complex aerial scenarios. Its precise monitoring abilities enable aerial equipment to perform tasks accurately at high speeds, enhancing robust QoS metrics.
GenAI employs a unique learning approach that maps latent distributions to key communication model parameters including path loss, delay, and angle measurements. This distributional perspective provides insights into correlated properties, enabling effective management of multiple parameters. However, current studies often rely on deterministic or stochastic optimizations, neglecting performance analysis under uncertainty conditions.




%% file: section5_new.tex
\section{MoE-transformer Actor Network for Diffusion-based Optimization}

In this section, we propose an MoE-transformer actor network to enhance the uncertainty-solving ability within the diffusion-based optimization framework.



\subsection{Diffusion-based Wireless Optimization}

\begin{figure*}[htp]
    \centering
    \includegraphics[width= 0.85\linewidth]{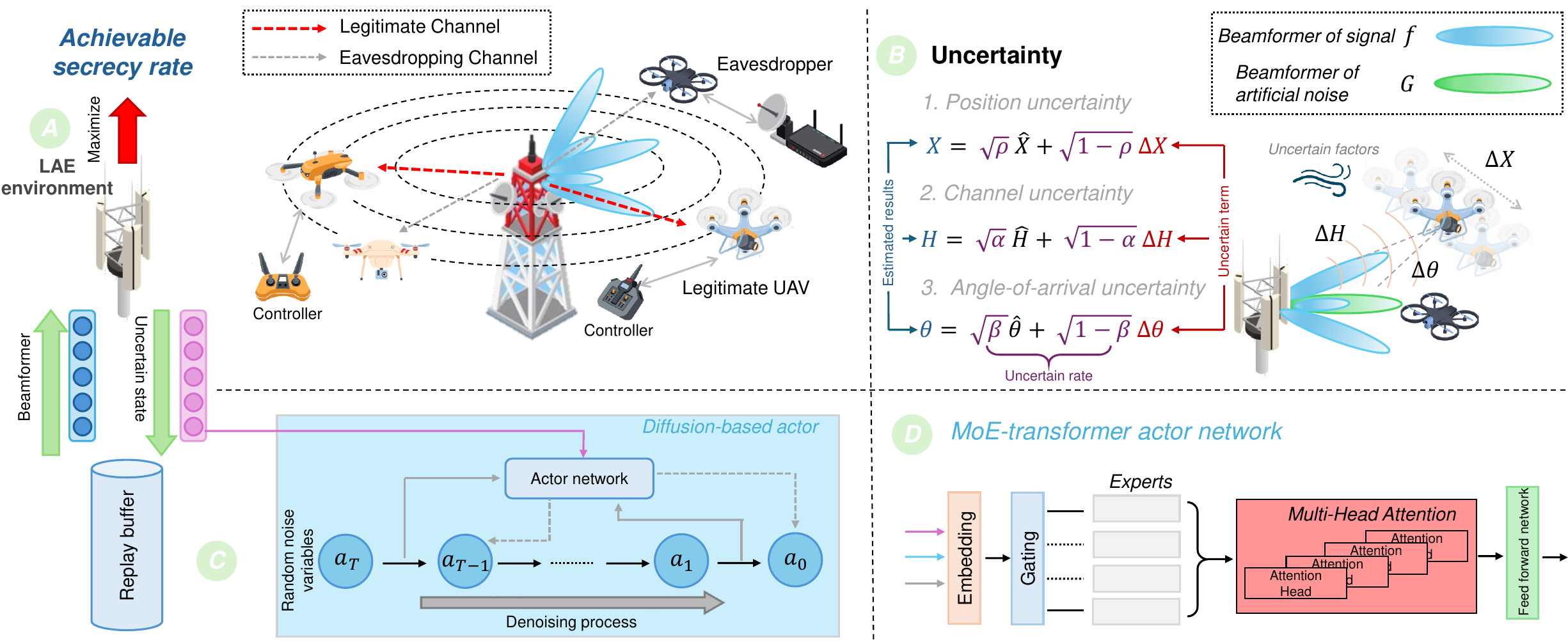}
    \caption{An illustration of GenAI-enabled robust beamforming for secure communication. \textit{Part A} presents the secure
communication architecture in LAENet.
    \textit{Part B} introduces the uncertainties in LAENets.
    \textit{Part C} presents the diffusion-based reinforcement learning framework.
    \textit{Part D} illustrates the network structure of the proposed MoE-transformer actor network.
    }
    \label{fig:case}
\end{figure*}

GenAI models, particularly diffusion models combined with reinforcement learning, have demonstrated significant advantages in wireless network optimization \cite{zhao2025enhancing}. These models inherently learn distributions, adapt to the effects of multi-scale fading in wireless channels, and exhibit a high level of adaptability that traditional algorithms lack.
For instance, the diffusion-based optimization framework GDM \cite{du2024enhancing} employs a diffusion model-based actor to generate actions, as illustrated in Fig. \ref{fig:case} \textit{Part C}. Through its multi-step denoising process, each step incrementally removes noise, enabling the actor to explore the entire state space and identify optimal actions. 
By contrast, other GenAI models, such as GAN-based or VAE actors, tend to either collapse to sub-optimal modes or struggle with the continuous, high-dimensional action spaces typical of beamforming and resource allocation problems \cite{sun2025generative}.
Moreover, diffusion models provide exact log likelihoods. Because GANs have no explicit density and VAE actors introduce approximation error, diffusion-based policies deliver more reliable uncertainty estimates and can adapt quickly to sudden channel variations~\cite{sun2025generative}.

However, the denoising optimization process faces several challenges, particularly in uncertainty optimization. 
The stochastic nature of diffusion model generation can lead to increased variability in actions under the same state, affecting network convergence. 
Therefore, our ongoing work augments the actor network within the diffusion-based optimization framework to achieve a balance between exploration and stability, further justifying diffusion models as a principled and practically effective choice for robust wireless optimization.

\subsection{MoE-transformer Actor Network}


Recently, Deepseek \cite{liu2024deepseek}, a generative LLM based on reinforcement learning, has emerged as a leading example of LLM.
It enhances model learning, generation capabilities, and convergence speed by integrating the MoE approach with the attention mechanism. 
\begin{itemize}
    \item \textbf{MoE}:  The MoE structure comprises a gating network and multiple expert networks. The gating network assigns scores to experts based on the input and activates them accordingly, allowing each expert to process data specially, leading to more precise inferences \cite{zhao2025enhancing}.
    In uncertainty optimization, the MoE can select the most suitable expert to estimate the uncertainty of the current state.
    
    \item \textbf{Attention}:  The attention mechanism dynamically assigns weights to different input elements, allowing the model to focus on the most relevant information. Computing attention scores selectively enhances important features, improving the efficiency of information processing. 
    Attention enables the model to capture uncertainty dependencies and contextual relationships, leading to more accurate predictions and representations.
\end{itemize}
In LAENets, the rapid variations in channel conditions, traffic loads, and adversarial threats significantly increase system uncertainties.
Multi-head attention mechanisms are particularly suitable for modeling fine-grained state dependencies and representing these uncertainties effectively on the fly.
An MoE layer retains this attention expressiveness while sparsely activating only the most relevant experts, slashing computing and energy demands to meet the tight resource availability of LAE edge nodes \cite{zhao2025enhancing}.
By integrating the two into an MoE-transformer network, the model effectively captures dynamic uncertainties while scaling computation adaptively, achieving the accuracy-efficiency trade-off essential for LAENets.


Inspired by this, we propose an MoE-transformer actor network structure to enhance the diffusion-based wireless optimization framework, as shown in Fig. \ref{fig:case} \textit{Part D}. Our approach integrates a transformer with multi-head attention to capture fine-grained dependencies and dynamic uncertainties across states and denoising steps, while MoE layers selectively activate the most relevant experts to stabilize generation and reduce computation.
The proposed actor network is specifically designed within the diffusion reinforcement learning framework to utilize the strengths of the MoE structure and attention mechanism, enabling more stable and robust action generation.
Therefore, the proposed method leverages the exploration and generative strengths of diffusion models to mitigate uncertainties in LAENets, enhancing overall system robustness.

\section{Case Study: GenAI-enabled Robust Beamforming for Secure Communication}

\subsection{Problem Description}

3D beamforming dynamically adapts beam directions and power allocations in response to environmental variations and UAV mobility, ensuring reliable and high-quality signal transmission. Robust beamforming leverages spatial diversity and directional signal control to mitigate eavesdropping risks and enhance confidentiality in LAENets \cite{dong2021deep}.

\subsection{Secure Beamformers}

We consider a secure communication scenario where a terrestrial BS transmits a confidential signal along with artificial noise to communicate with legitimate UAVs and evade eavesdropping, as depicted in Fig. \ref{fig:case} \textit{Part A}.
The terrestrial BS employs a uniform planar array (UPA) with dimensions \(N_x \times N_y\), whereas both legitimate UAVs and eavesdroppers (Eves) are equipped with uniform linear arrays (ULAs) comprising \(N_b\) and \(N_e\)
antennas, respectively.
Moreover, environmental factors, such as wind speed, will introduce estimation errors in the target UAV's location. The covert nature of the eavesdropper further exacerbates uncertainties, while errors in channel estimation and angle-of-arrival estimation also influence beamforming accuracy. 
As illustrated in Fig. \ref{fig:case} \textit{Part B}, these uncertainties are incorporated in the optimization input. In simulations, we emulate varying uncertainty levels by injecting controlled noise.

\subsection{Optimization Design}

To ensure reliable communication while mitigating eavesdropping risks, we accomplish secure beamforming by solving an optimization problem. Our objective is to maximize the achievable secrecy rate (ASR) by optimizing the beamformers for both the confidential signal and artificial noise, accounting for various sources of uncertainties. To restrict the eavesdropper’s ability, we impose a constraint that limits the capacity of the eavesdropping channel to remain below a specific threshold $C_{eve}$.
We designed a reward function with a penalty item to ensure reinforcement learning algorithms can learn the optimal beamforming strategy under the constraints. Specifically, the reward function is the ASR minus the capacity of the eavesdropping channel over the threshold $C_{eve}$. 

To comprehensively evaluate the proposed algorithm's effectiveness in enhancing robustness, we solve the secure beamforming via varying robustness levels of optimization.
\begin{itemize}
    \item \textbf{Stochastic Optimization} aims to maximize the mathematical expectation ASR for a given beamforming action under uncertainties. 
    We incorporate the eavesdropping channel capacity as a penalty term to enforce constraints.
    
    \item \textbf{Chance-Constrained Optimization} maximizes ASR while ensuring constraint satisfaction with a certain probability $P_{eve}$.
    Since direct probability computation is intractable, 
    we employ a Monte Carlo method to approximate constraint satisfaction probability for all reinforcement learning methods. 
    When the constraint satisfaction probability is greater than $P_{eve}$, we only calculate the reward brought by ASR to maximize ASR from a probability perspective.
    
    \item \textbf{Robust Optimization} focuses on guaranteeing worst-case performance under uncertainties.
    For a given beamforming action, we use the Monte Carlo method to approximate 
    the worst-case with 
    the lowest achievable reward. Then, we maximize this lower bound to enhance the robustness of the generated beamforming.
\end{itemize}

\begin{figure*}[t]
\centering
\begin{subfigure}{.24\textwidth}
  \centering
  \includegraphics[width=1.1\linewidth]{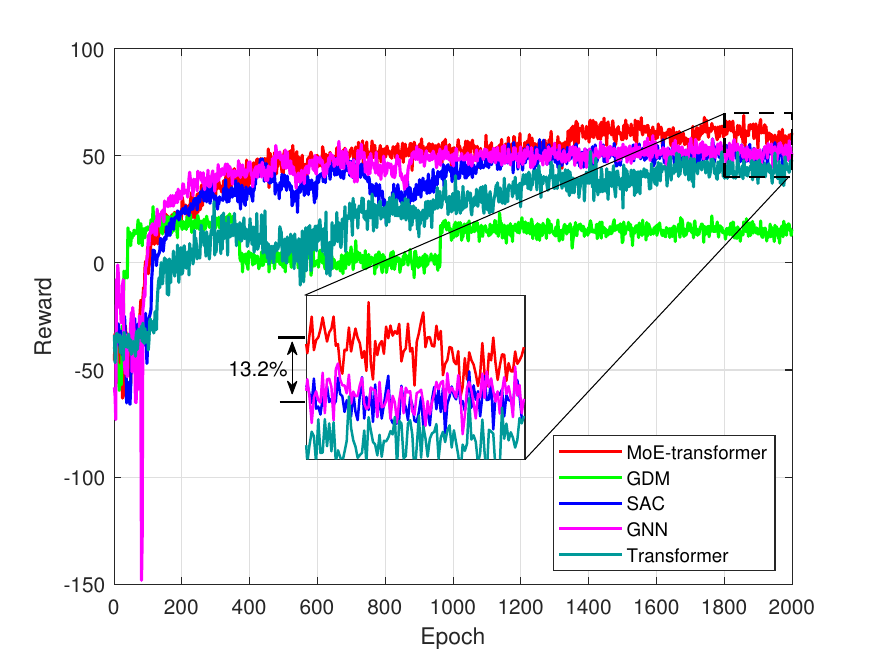}
  \caption{Stochastic}
  \label{fig:sub1}
\end{subfigure}
\begin{subfigure}{.24\textwidth}
  \centering
  \includegraphics[width=1.1\linewidth]{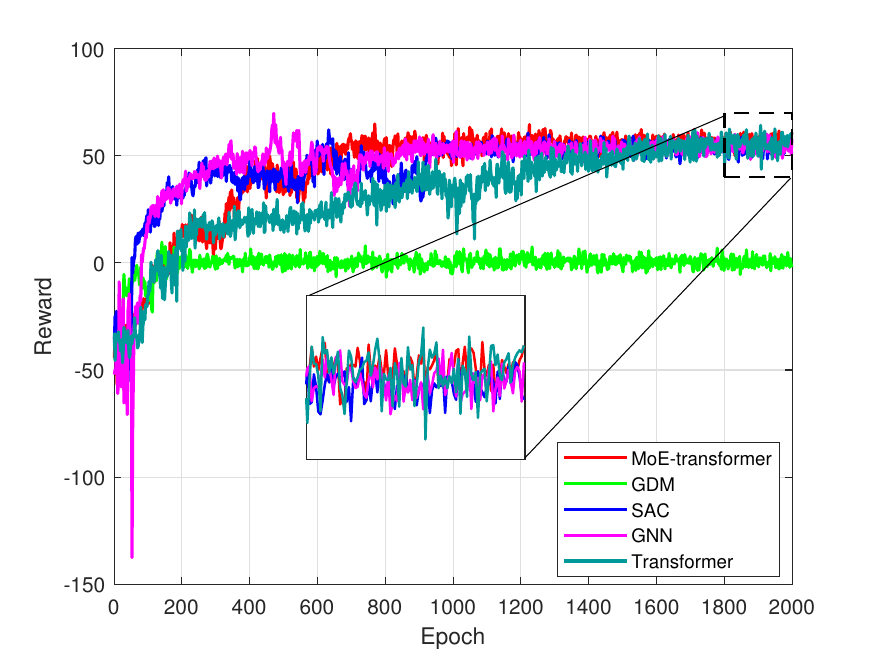} 
  \caption{Chance-constrained}
  \label{fig:sub2}
\end{subfigure}%
\begin{subfigure}{.24\textwidth}
  \centering
  \includegraphics[width=1.1\linewidth]{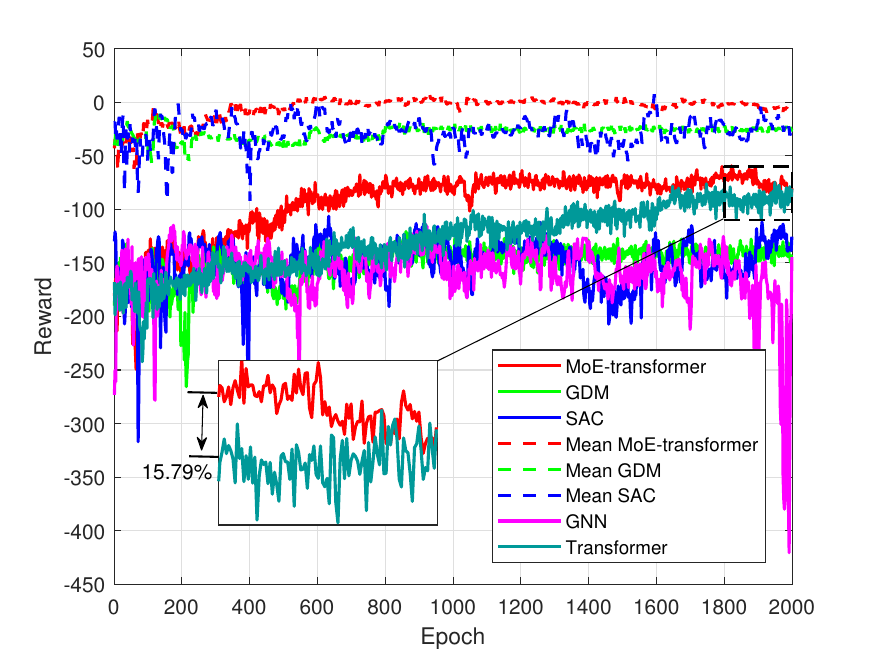}
  \caption{Robust}
  \label{fig:sub3}
\end{subfigure}
\begin{subfigure}{.24\textwidth}
  \centering
  \includegraphics[width=1.0\linewidth]{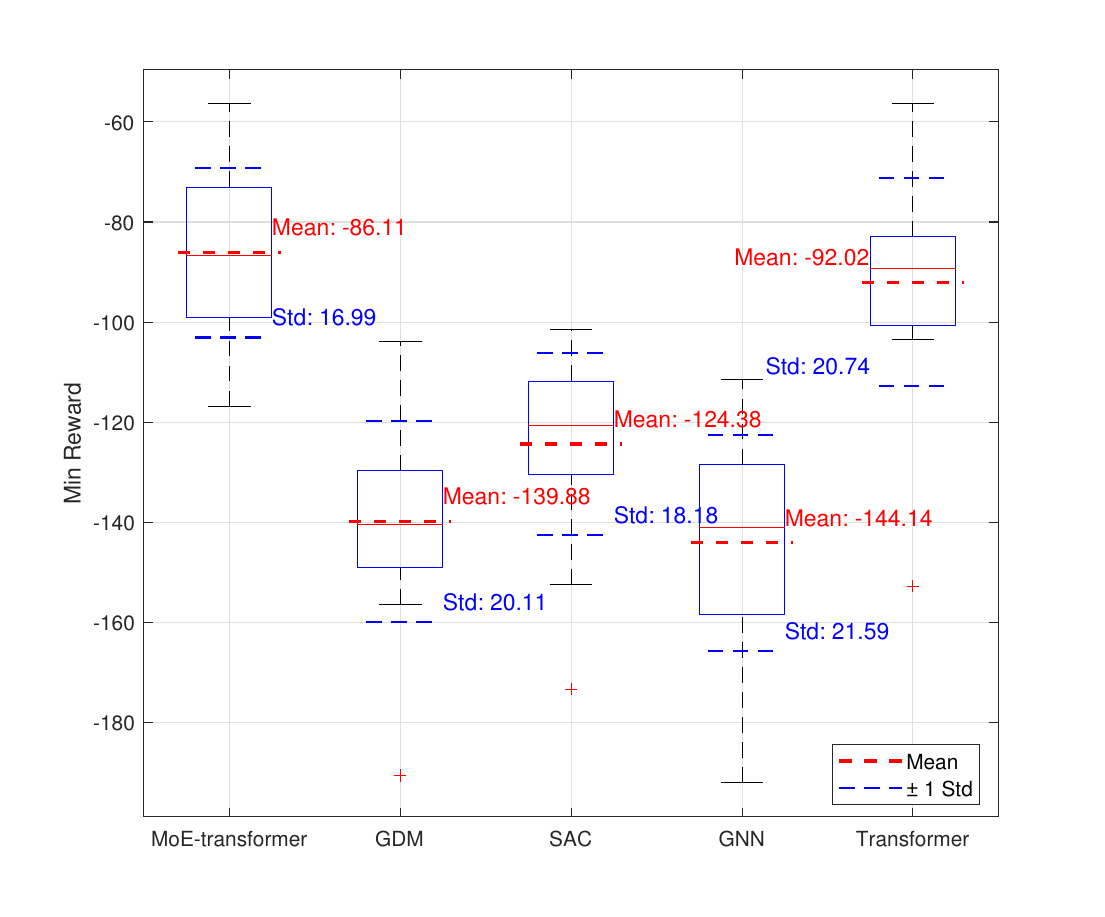}
  \caption{Inference Reward}
  \label{fig:sub4}
\end{subfigure}
\caption{Numerical results of the proposed algorithm and baselines.
The server is equipped with an Ubuntu 22.04 operating system and powered by an Intel(R) Xeon(R) Silver 4410Y 12-core processor and an NVIDIA RTX A6000 GPU.
}
\label{fig:test}
\end{figure*}

\subsection{Numerical Results}

In the experiments, we consider a terrestrial BS with a \(4\times 4\) UPA, a legitimate UAV, and an eavesdropper, each equipped with $N_b=N_e=6$ antennas. 
The beamforming design accounts for both position and channel uncertainties while adopting the same channel power gain settings as in \cite{dong2021deep}, and $C_{eve}= 3 bps/Hz$.
In chance-constrained optimization, we set the constraint probability $P_{eve}=70\%$. 
We evaluate our method by benchmarking it against the non-GenAI method SAC \cite{wang2022deep}, GNN \cite{tang2025deep}, transformer \cite{sun2025generative}, and the GenAI method GDM \cite{du2024enhancing}, which does not use an MoE-transformer Actor Network.
All networks are trained for 2000 epochs with a learning rate of \(1 \times 10^{-4}\).
Our MoE-transformer actor network\footnote{For more implementation details, please refer to the code available at https://github.com/ChangyuanZhao/LAENet\_robust\_opt.} employs 4 experts with top-2 selection, uses 4 attention heads, and adopts a hidden dimension of 256. In the GDM setting, we configure the diffusion process with 6 steps and adopt the same network architecture as in \cite{du2024enhancing}.


The learning curves for the three optimizations are presented in Figs. \ref{fig:sub1}, \ref{fig:sub2}, and \ref{fig:sub3}, where the red, green, blue, purple, and cyan curves represent the MoE-Transformer, GDM, SAC, GNN, and transformer algorithms, respectively. 
Moreover, the dashed lines in Fig. \ref{fig:sub3} are the average ASR reward under robust optimization.
The results show that our proposed algorithm consistently demonstrates the best learning capability across all optimization settings. In stochastic optimization, it moves beyond SAC and GNN by about 13\%.  
Compared to GDM algorithm, whose learning curve remains noisy and slow to settle, our proposed algorithm achieves over 200\% improvement. In chance-constrained optimization, most algorithms exhibit similar convergence behavior to our proposed method, which maintains a relatively fast convergence speed.
In robust optimization, where all returns shift into negative territory, the MoE-transformer remains the least degraded. Its reward sits around 15.8\% above transformer and about 44\% higher than SAC. Collectively, these curves show that pairing a sparse mixture-of-experts backbone with diffusion-guided policy learning not only accelerates convergence but also provides consistently superior robustness across the full spectrum of optimization objectives.
Although the MoE-Transformer activates 4 experts, its per-iteration wall-clock time remains 0.01906 s, versus 0.01813 s for the transformer, 0.01625 s for SAC and GDM, and 0.01875 s for GNN.
This around 15\% latency gap is far below the 100 ms control-loop budget typical for LAE edge nodes. 
Consequently, the proposed architecture satisfies real-time and energy-efficiency requirements while still delivering the robustness gains reported above.

Finally, we compared the inference reward of all methods in robust optimization as shown in Fig. \ref{fig:sub4}.
The proposed MoE-transformer method achieves the lowest negative minimum reward with the lowest variance of 16.99, indicating significantly steadier behavior. In contrast, GDM, SAC, GNN, and the transformer model all show larger variances, indicating that their worst-case returns fluctuate more significantly, and occasionally result in extremely low returns. The less variance seen in the MoE-transformer's box plot results in a more dependable safety margin against adverse conditions, with more robust beamforming policies.


\subsection{Open Challenges}

Despite the encouraging gains shown in Figs. \ref{fig:sub1}–\ref{fig:sub4}, our study also uncovers several open challenges that must be tackled before GenAI-enabled robust beamforming can be translated from the lab to large-scale LAENets.
\begin{itemize}
    \item \textbf{Data requirement:} GenAI needs to learn the policy distributions. Greater uncertainties directly translate to a larger data requirement before the optimal LAENet strategy can be identified.
    \item \textbf{Hyper-parameter sensitivity:} Performance depends on the noise schedule, denoising depth, and MoE expert count. Field deployments will demand continual re-tuning to balance latency, energy, and robustness.
    \item \textbf{Monte-Carlo simulation limitation:} Current solutions justify reliability with empirical simulations only. Without formal worst-case guarantees, true end-to-end robustness remains unproven for safety-critical aerial links.
\end{itemize}





%% file: con.tex
\section{Future Directions}

\subsection{Adaptive GenAI for Dynamic Environments}

LAENets, such as drone-based logistics and aerial monitoring systems, operate in highly dynamic and unpredictable environments. Customizing GenAI to adapt to real-time changes in network topology, environmental conditions, and mission requirements is essential. Techniques, including online learning, real-time model adaptation, and context-aware optimization can be explored to ensure that GenAI remains robust and effective in these ever-changing scenarios.

\subsection{GenAI-enabled Anomaly Detection and Recovery}

Ensuring reliability in LAENets requires robust mechanisms for detecting and recovering from anomalies such as signal interference, equipment malfunctions, and unexpected obstacles. GenAI can be utilized to develop advanced anomaly detection frameworks that integrate multimodal data, including sensor, visual, and communication data. Furthermore, designing GenAI models that suggest real-time recovery strategies, such as dynamic path adjustments or alternate resource allocations, can significantly enhance network resilience.

\subsection{Energy-Aware GenAI Optimization}

Energy efficiency is a critical constraint in LAENets, where nodes often rely on limited power sources such as drone batteries. Future research should focus on developing energy-aware GenAI models that balance computational demands with power constraints. Techniques such as lightweight model architectures, energy-efficient inference, and collaborative processing across network nodes can help extend operational lifespans while maintaining the robustness of the network.

\section{Conclusion}

In this paper, we have explored the application of GenAI techniques to enhance the robustness of LAENets. Initially, we conducted a systematic analysis of the robustness requirements for LAENets, followed by a comprehensive review of robust QoS metrics from the perspective of the wireless physical layer.
Subsequently, we have performed an in-depth analysis of existing GenAI-enabled approaches for robustness enhancement and their potential applications. Building on this, we have proposed a diffusion-based reinforcement framework with an MoE-transformer actor network.
In the robust beamforming case study, the proposed framework has effectively learned the beam pattern with various uncertainties, achieving an improvement of approximately 13\% ASR in stochastic optimization,
and around 15\% ASR in robust optimization than four learning baselines.
These results underscore the critical role of GenAI in enhancing the robustness of LAENets and highlight the need for further exploration of its applications.